\RequirePackage{amsmath}
\documentclass[12pt,a4paper]{iopart}
\usepackage{iopams,verbatim,units,bbm,color,multirow,graphicx,tabularx}

\usepackage[colorlinks]{hyperref} 

\definecolor{iblue}{rgb}{0.07,.24,.67}
\definecolor{sred}{rgb}{0.97,0,.07}

\newcommand{\ket}[1]{\left | #1 \right \rangle}

\newcommand{\beq}{\begin{equation}}
\newcommand{\eeq}{\end{equation}}
\newcommand{\beqa}{\begin{eqnarray}}
\newcommand{\eeqa}{\end{eqnarray}}

\newcommand{\spinhalf}{spin-$\nicefrac 12$}
\renewcommand{\u}{{\widehat{u}}}
\renewcommand{\v}{{\widehat{v}}}
\newcommand{\x}{{\widehat{x}}}
\newcommand{\y}{{\widehat{y}}}
\newcommand{\z}{{\widehat{z}}}
\newcommand{\m}{{\widehat{m}}}

\begin{document}
\title{Holonomic quantum computing in symmetry-protected ground states of spin chains}

\author{Joseph M. Renes$^{1,2}$, Akimasa Miyake$^{3,4}$, Gavin K. Brennen$^5$, and Stephen D. Bartlett$^6$}
\address{$^1$ Institute for Applied Physics, Technische Universit{\"a}t Darmstadt, Hochschulstr.~4a, 64289 Darmstadt, Germany}
\address{$^2$ Institute for Theoretical Physics, ETH Zurich, Wolfgang-Pauli-Str.\ 27, 8093 Z\"urich, Switzerland}
\address{$^3$ Perimeter Institute for Theoretical Physics, 31 Caroline Street North,
Waterloo Ontario, N2L 2Y5, Canada}
\address{$^4$ Center for Quantum Information and Control, University of New Mexico, Albuquerque~NM 87131, USA}
\address{$^5$ Centre for Engineered Quantum Systems, Macquarie University, Sydney, NSW~2109, Australia}
\address{$^6$ Centre for Engineered Quantum Systems, School of Physics, University of Sydney, NSW 2006, Australia}

\date{\today} 

\begin{abstract}
While solid-state devices offer naturally reliable hardware for modern classical computers, thus far quantum information processors resemble vacuum tube computers in being neither reliable nor scalable. Strongly correlated many body states stabilized in topologically ordered matter offer the possibility of naturally fault tolerant computing, but are both challenging to engineer and coherently control and cannot be easily adapted to different physical platforms.
We propose an architecture which achieves some of the robustness properties of topological models but with a drastically simpler construction. Quantum information is stored in the symmetry-protected degenerate ground states of spin-1 chains, while quantum gates are performed by adiabatic non-Abelian holonomies using only single-site fields and nearest-neighbor couplings. Gate operations respect the symmetry, and so inherit some protection from noise and disorder from the symmetry-protected ground states. 
\end{abstract}
\maketitle
\tableofcontents

\section{Introduction}

Research into quantum computing has produced an ever increasing variety of physical realizations of a quantum bit (qubit) that aim to simplify the construction of a practical quantum computer. The original quantum circuit model, where the logic of the quantum computer is implemented by the {\it time-dependent} control of the system Hamiltonian, is a common choice in practical realizations. Despite its theoretical simplicity, the model makes significant practical demands.
For example, the controlling fields have to be sharp and precise enough (in both space and time) to 
execute a universal set of elementary gates with very high fidelity, while avoiding undesired decoherence 
resulting from the inevitable coupling of the system to the environment.  For error rates below a certain (noise model dependent) threshold, quantum error correction can be used to allow for scalable fault tolerant quantum computation~\cite{gottesman_introduction_2009}.  To date, however, no physical architectures meet the advertised error thresholds, and while there exist theoretical techniques to further reduce errors, such as dynamical decoupling~\cite{west_high_2010}, it is sensible to consider how the challenge of fault-tolerance could be simplified by a more elaborate architecture.

\begin{figure*}[hbtp!]
\begin{center}
\includegraphics[width=\textwidth]{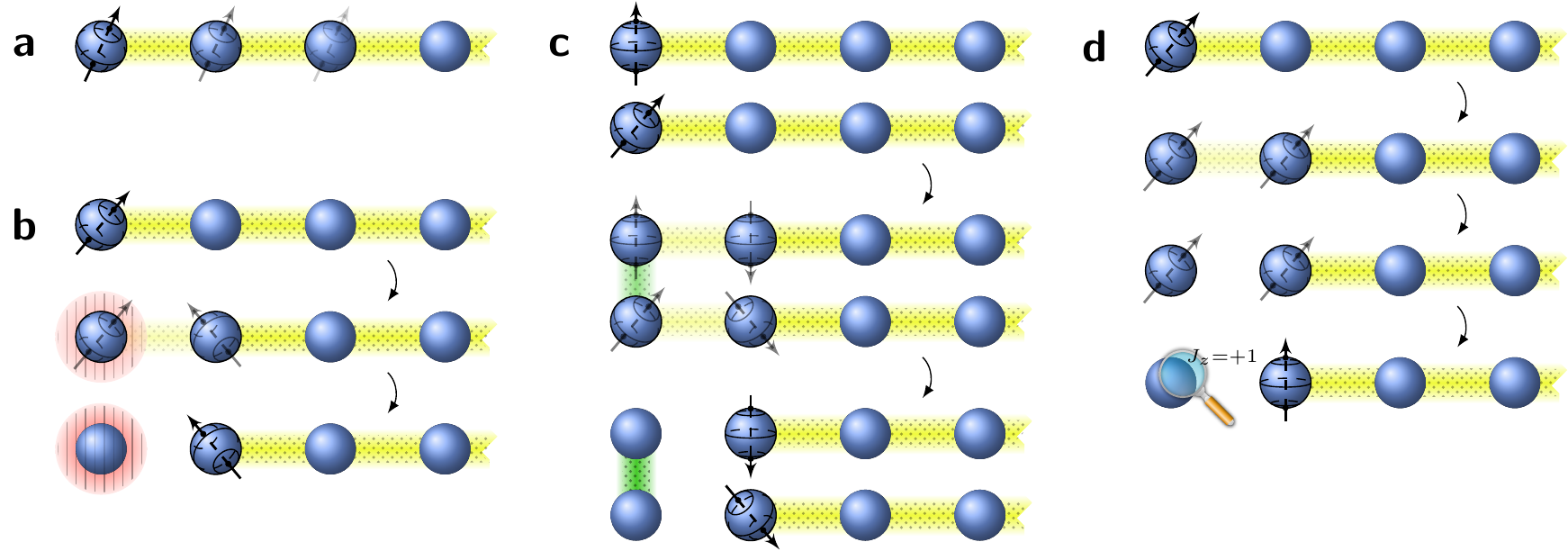}
\end{center}
\vspace{-2mm}
\caption{
\label{fig:chains}
Spin-chain qubit encoding and logical operations. {\bf \textsf a} In the Haldane phase, ground states of spin-1 chains correspond to edge-localized \spinhalf\ degrees of freedom and are used as encoded qubits. Blue spheres represent the spin-1 objects, and the yellow bands their coupling. The encoded qubit is represented as an arrow piercing the Bloch sphere, and its localization by Bloch spheres at successive spin-1 elements on the end of the chain. {\bf \textsf b} Adiabatically decoupling the boundary spin from its immediate neighbor while simultaneously turning on a local field, shown in red, realizes a single qubit operation, transferring the encoded qubit to the slightly shorter chain while effecting a $\pi$ rotation about the local field axis. The time-reversed process works similarly, and combining $\pi$ rotations around different axes enables the execution of any single-qubit operation. {\bf \textsf c} Appropriate adiabatic coupling of boundary spins of neighboring chains (shown in green) and simultaneous decoupling from their respective chains realizes a {\sc cphase} gate followed by a joint $\pi$ rotation about the $\hat{x}$ axis (pointing out of the page).
{\bf \textsf d} Measurement and initialization of the encoded qubit can be performed as in~\cite{miyake_quantum_2010}.  The coupling to the boundary spin is adiabatically switched off and subsequently a measurement made of the boundary spin in the basis $\ket{S_z=m}$. The result $m=1$ ($m=-1$) corresponds to a projection of the qubit onto $\ket{0}$ ($\ket{1}$), while $m=0$ corresponds to a $\hat{z}$-axis $\pi$ rotation. The rotation can be undone by recoupling the boundary spin as in {\bf \textsf a}, and so the readout operation can be repeated until a nonzero $m$ outcome is obtained.
}
\vspace{-5mm}
\end{figure*}

Our architecture, summarized schematically in Figure~\ref{fig:chains} and explained in Sec.~\ref{Architecture}, 
combines two key ideas to advance this goal while making modest technological demands. The first is to use a non-Abelian holonomy (geometric phase)~\cite{wilczek_geometric_1989} of an adiabatic transformation for the logical gate action~\cite{zanardi_holonomic_1999,sjoeqvist_new_2008}, because such a geometrical quantity is more robust to temporal inaccuracy of the controlling fields than dynamically generated transformations.   This mechanism is doubly advantageous when the degenerate subspace relevant for the non-Abelian holonomy is  physically well-isolated  from the other degrees of freedom in the Hilbert space, as this reduces the chance of leakage errors. To this end, the second idea is to define our qubit as the gapped ground subspace of a two-body interacting system of small numbers of spin particles (which we require to be integer spins, e.g., spin 1's, for our purpose).    By basing our strongly-interacting system on a model possessing symmetry-protected topological order (SPTO)~\cite{gu_tensor-entanglement-filtering_2009,pollmann_entanglement_2010}, the degeneracy of this ground space is \emph{protected} by a symmetry and cannot be broken by any symmetry-respecting perturbations.  Our logic gates respect this symmetry, and therefore will preserve the encoding.  In addition, the energy gap in our model provides built-in protection against some errors~\cite{brennen_measurement-based_2008}.  In our proposed architecture, a constant energy gap guarantees that a sufficiently slow adiabatic transformation also preserves the ground state encoding.  

The interplay between strongly-correlated ground states of spin lattices and quantum computation has been the subject of considerable recent activity e.g., in~\cite{brennen_measurement-based_2008,barrett_transitions_2009,doherty_identifying_2009,skrvseth_phase_2009,miyake_quantum_2010,bartlett_quantum_2010,miyake_quantum_2010-1,else_quantum_2012,darmawan_AKLT_2012}.  We emphasise that our proposed implementation is distinct from proposals for topological quantum computation, which is essentially the implementation of a quantum circuit model using exotic non-Abelian statistics of quasi-particles anyons in topological ordered systems~\cite{nayak_non-abelian_2008}, including 1-D chains~\cite{alicea_non-abelian_2011}.  Our architecture is not topologically protected against arbitrary local perturbations, instead being protected by the dihedral group $D_2$ (equivalently $\mathbb{Z}_2 \times \mathbb{Z}_2$) of $\pi$ rotations  of the entire chain  about two orthogonal spatial directions.  
Such SPTO systems do not offer the same robustness to arbitrary finite-range perturbations as do topologically ordered systems, but they do guarantee both that the logical subspace is protected by the energy gap and also that its degeneracy is preserved under $D_2$-invariant perturbations.

The $D_2$ group of symmetry operations also serves to define the logical Pauli operators of the encoded qubits in our scheme,
which can be interpreted as acting on the so-called \textit{edge state}:  a pair of fractionalized emergent \spinhalf\ degrees of freedom, one at each end of the spin chain~\cite{kennedy_exact_1990}.  These edge states were speculated to be useful for quantum information processing in Ref.~\cite{katsura_exact_2007}.   A key advantage of encoding in these edge states is that their existence and degeneracy is a robust property of an entire symmetry-protected phase, i.e., it does not rely on fine-tuning of the Hamiltonian.  Such encodings have been used in measurement-based models~\cite{brennen_measurement-based_2008,miyake_quantum_2010}; in contrast, our qubit operations  are performed by adiabatic manipulation of the edge fields and couplings, with no need to ``consume'' the spins of the chain by measuring them. Our proposal is essentially an adiabatic version of~\cite{miyake_quantum_2010}, echoing the adiabatic replacement of gate teleportation~\cite{bacon_adiabatic_2009} and the cluster state architecture~\cite{bacon_adiabatic_2010}.\footnote{In fact, in one-dimension the cluster state possesses $\mathbb{Z}_2 \times \mathbb{Z}_2$ SPTO~\cite{son_quantum_2011,else_quantum_2012} and,
in the framework of~\cite{chen_classification_2011}, it can be transformed to the same fixed-point state which lies in the Haldane phase as we describe here, in terms of local unitaries~\cite{verstraete_renormalization-group_2005} chosen to be symmetry-respecting.}

Symmetries of the phase also play a crucial role in defining logical gates. Single qubit gates, the most basic of which in this proposal are $\pi$ rotations about a chosen spatial direction, rely on the system possessing $D_2$ symmetry of $\pi$ rotations containing the rotation axis. The symmetry basis can be adapted from gate to gate, meaning full rotational invariance is not required in order to perform rotations about arbitrary axes. The two-qubit \textsc{cphase} gate relies on a different symmetry, one generated by spatial reflection in two orthogonal planes as well as combined rotations of spins in both chains. This symmetry group is not analytically known to protect the Haldane phase, but numerical investigation shows that the ground state degeneracy is not lifted during the gate action. 

The outline of the paper is as follows. In Sec.~\ref{Architecture} we give a detailed account of the quantum computation protocol and show how the holonomic gates work on qubits encoded in the degenerate ground state of a SPTO spin chain.  In Sec.~\ref{NoiseAnalysis} we analyse the effect of various noise sources in our system and describe how the information processing can be made fault tolerant by using error correction and cooling on top of our improved coherent operations.  We conclude with a summary of the results in Sec.~\ref{Sum}.

\section{Architecture}
\label{Architecture}

Despite its non-standard appearance, computation in our model (see Fig.~\ref{fig:chains}) proceeds just as in the original circuit model,  with gates based on adiabatic transformations of degenerate energy eigenstates, and thus it can be thought of as an instance of holonomic quantum computation~\cite{zanardi_holonomic_1999,sjoeqvist_new_2008}. Accordingly, the methods of fault-tolerant holonomic computation~\cite{oreshkov_holonomic_2009,oreshkov_fault-tolerant_2009} can be applied. 


In our model, each qubit is encoded in an edge mode associated with the (near-) degenerate ground state of a spin-1 chain in the Haldane phase.  The Haldane phase of a spin-1 chain is a  gapped phase,  possessing SPTO that can be characterised (among several possible symmetries) by the symmetry $D_2$~\cite{gu_tensor-entanglement-filtering_2009,pollmann_entanglement_2010,chen_classification_2011}.   Unlike phases characterized by the traditional Landau's spontaneous symmetry breaking, this ground state still respects the symmetry and is now considered a 1D counterpart of the ``topological'' phase because of the following notable properties:  (1) a four-fold degenerate gapped ground state in the thermodynamic limit, with finite chains possessing an exponentially-small splitting of this degeneracy in terms of the chain length; (2) a corresponding degeneracy of the entanglement spectrum of the ground state for any bipartition~\cite{pollmann_entanglement_2010}; (3) fractionalised spin-1/2 degrees of freedom on the boundaries of a finite chain -- the \emph{edge states}.   Canonical points within this phase are the spin-1 Heisenberg antiferromagnet, as well as the Affleck-Kennedy-Lieb-Tasaki (AKLT) model~\cite{affleck_valence_1988}.   The ground state(s) of the AKLT model possess an exact valence-bond solid (VBS) description with bond dimension 2, and exact four-fold degeneracy even for finite chains.  This exact description of the AKLT ground state allowed for initial demonstrations that it was useful as a resource for measurement-based quantum computation~\cite{brennen_measurement-based_2008}, and subsequently it was realised that many of its quantum-computational properties can be extended throughout the phase~\cite{bartlett_quantum_2010,miyake_quantum_2010-1,else_quantum_2012}.

 We now consider quantum gates on this encoding.  A  universal set of quantum gates can be realized using the two-body interaction of the \textsc{cphase} gate and a handful of local field settings for single-qubit logic, along with turning off the coupling of boundary spins and their neighbors. Critically, only two-body couplings are involved, a feature absent in previous proposals of adiabatic holonomically-controlled architectures~\cite{oreshkov_holonomic_2009, oreshkov_fault-tolerant_2009, bacon_adiabatic_2009,bacon_adiabatic_2010}. (In these prior proposals, the necessity of more than 2-body interactions is overcome by the use of perturbation gadgets, but the fact that ideal gate operation is only achieved in the limit of zero perturbation implies a very delicate control of the energy scale in practice.) Finally, initialization and readout can be performed by adapting the scheme of Ref.~\cite{miyake_quantum_2010}.  

\begin{figure}[h]
\begin{center}
\includegraphics[width=6cm]{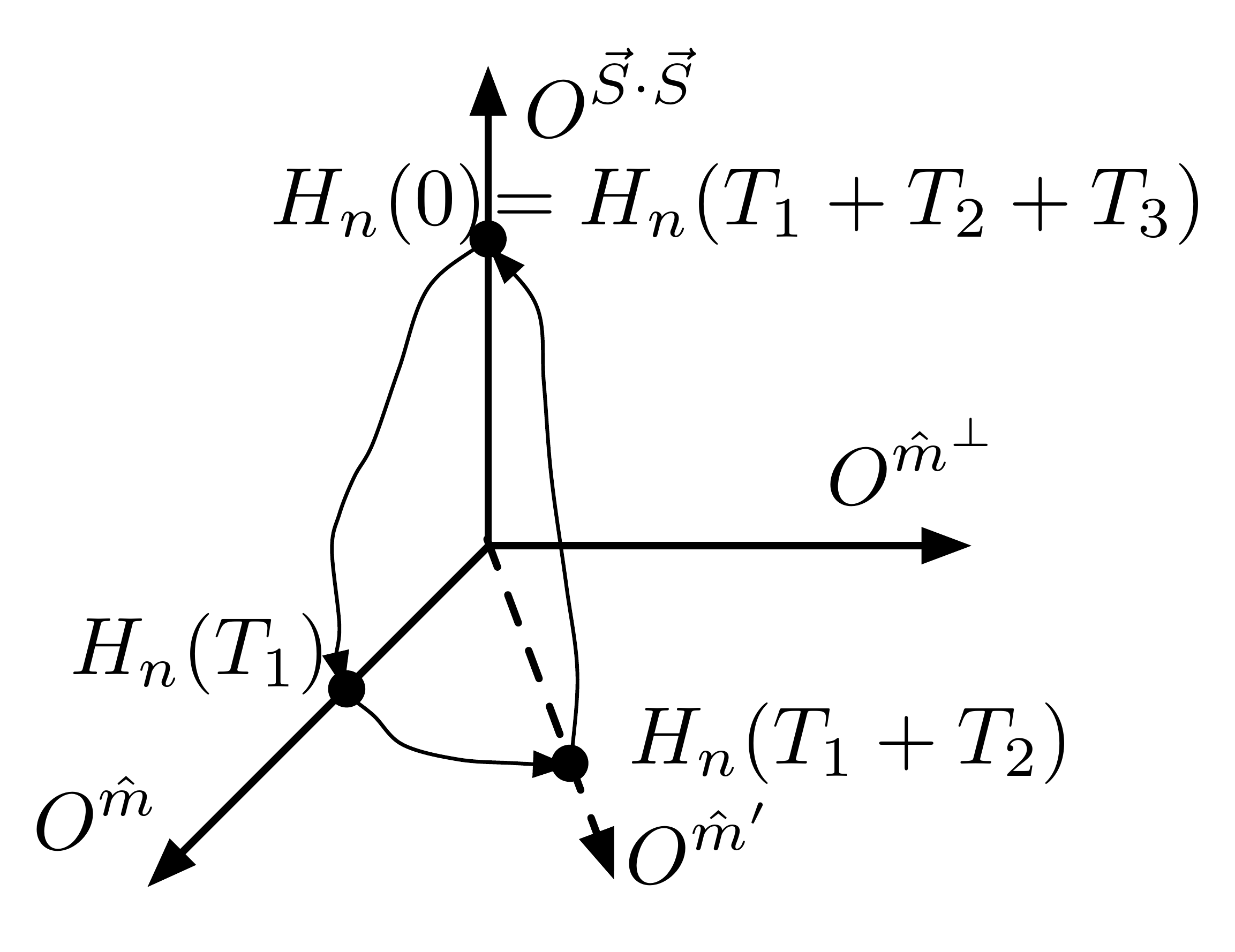}
\caption{\label{fig:holo} Sketch of a holonomic path in parameter space of the chain Hamiltonian $H_n(t)$, traced out by time dependent coupling between the boundary spin$-1$ and its neighbor, which realizes a single qubit rotation.   The axes parameterize weights on the operators $O^{\vec{S}\cdot\vec{S}}=\vec{S}_1\cdot\vec{S}_2,$ $O^{\widehat{r}}=(S_1^{\widehat{r}})^2-\frac{\bf 1}{3}$ (where $\widehat{r}$ are unit vectors in $\mathbb{R}^3$) on $H_n(t)$.  The operators $\{O^{\vec{S}\cdot\vec{S}},O^{\m},O^{\m^{\perp}}\}$  constitute a trace orthogonal set  and are all $D_2$-invariant.  The path consists of three adiabatic steps which each take a time $T_j\sim 1/\Delta$. The qubit states are maintained in the ground state for paths that take place in the positive octant of the parameter space.  The holonomy is a qubit rotation about the axis $\m\times\m'$ by an angle equal to twice angle formed by $\m$ and $\m'$.
}
\end{center}
\end{figure}

\subsection{Holonomic Gates}

Consider a chain of $n$ spin-1 particles, each coupled to its nearest neighbors via the Heisenberg coupling with Hamiltonian
\begin{align}
\label{eq:Ham}
H_{n}^{\rm open}=J\sum_{j=1}^{n-1}\vec{S}_j\,{\cdot}\,\vec{S}_{j{+}1},
\end{align}
for $J>0$. This Hamiltonian, which is $D_2$-invariant and gapped, describes a Haldane phase that is distinct from the trivial phase in that local perturbations that respect the $D_2$ symmetry cannot connect these phases without closing the gap~\cite{gu_tensor-entanglement-filtering_2009,pollmann_entanglement_2010}. The ground state of such open chains is nearly fourfold degenerate, with a splitting that decays exponentially in the chain length. In particular, the singlet (triplet) states are ground states for even (odd) length $n$ chains but the splitting scales like $O((-1)^n e^{-n/\xi}n^{-1/2})$ where $\xi\approx 6.03$ is the correlation length, whereas the gap to bulk spin-$2$ excitations is $\Delta\approx 0.41 J$~\cite{hagiwara_observation_1990,white_density_1992}.

The four ground states correspond to two fractionalized \spinhalf\ degrees of freedom, one at each edge~\cite{kennedy_exact_1990,hagiwara_observation_1990,polizzi_s1/2_1998}. 
In order to model the state of one edge only, say the left, we ``terminate'' the right end of the chain with an additional \spinhalf\ particle (which may be fictitious), coupling the termination \spinhalf\ to the boundary spin-1 with the Heisenberg interaction so that the total Hamiltonian is 
\begin{align}
\label{eq:Hamterm}
H_{n}^{\rm term}=J\sum_{j=1}^{n-1}\vec{S}_j\,{\cdot}\,\vec{S}_{j{+}1}+J\vec{S}_n\cdot\vec{s}_{n+1},
\end{align}
This effectively fixes boundary conditions on the right edge (in physics language), or purifies the right edge mode (in the language of quantum information theory), leaving a twofold degenerate ground state.\footnote{The termination coupling is also $D_2$-invariant, yet \emph{does} break the degeneracy of the ground states of the open chain, in apparent contradiction to the claim of degeneracy protection. Strictly speaking, the degeneracy is only protected for $D_2$-invariant perturbations which do not alter the representation by which the ground state transforms~\cite{chen_classification_2011}.  By coupling to an additional \spinhalf\ system rather than an integer spin, this representation becomes projective.   (Recall that a (true) representation $T$ of a group $G$ must satisfy $T(g_1)T(g_2)=T(g_1g_2)$ for all $g_1,g_2\in G$.  A projective representation need only satisfy the weaker condition that $T(g_1)T(g_2)=\omega(g_1,g_2)T(g_1g_2)$ for some phase $\omega$.)} Though presented here as merely a mathematical device, the extra \spinhalf\ system could be realized as part of the system. 

Several facts justify this model of the edge states. First, at the AKLT point the description is exact, as ground states of the terminated Hamiltonian are also ground states of the unterminated Hamiltonian due to the frustration-free property of the valence-bond solid AKLT ground state~\cite{affleck_valence_1988}. Second, away from AKLT, numerical results show that for chains of modest length, measuring the termination spin of a $H_n^{\rm term}$ ground state results in a high-fidelity approximation to a $H_n^{\rm open}$ ground state. Specifically, fidelities at the Heisenberg point exceed 0.998 for chains of length up to 12, decreasing roughly linearly with chain length. Finally, direct numerical simulation of the gates and measurements presented below make clear that the scheme works well even for quite short chains: single qubit gate fidelities at the Heisenberg point exceed 0.999 even for chains of length as short as six.


The remaining fractionalized edge degree of freedom can be used to encode a single qubit, with logical Pauli operators $\bar{\sigma}^{\widehat{m}}=\Sigma^{\widehat{m}}_n$ for the length-$n$ chain defined as 
global rotations around the $\widehat{m}$ axis: 
\begin{align}
   i  \Sigma^{\widehat{m}}_n=\Big(\bigotimes_{j=1}^n \exp({i\pi S^{\m}_j})\Big)\otimes \exp\left(i\tfrac{\pi}{2}\sigma^{\m}\right)
\end{align}
As these operators generate a projective representation of the $D_2$ symmetry (the Pauli group), this qubit encoding is well-defined throughout the phase.

The encoded qubit can be manipulated by adiabatically weakening the boundary spin coupling and turning on a local term, as in the Hamiltonian
\begin{align}
\label{eq:singlequbitdyn}
H_{n}(t)=f(t)J(S^{\widehat{z}}_1)^2+g(t)J\vec{S}_1\,{\cdot}\,\vec{S}_{2}+H_{n{-}1}^{\rm term},
\end{align}
with monotonic $f,g$ obeying $f(0){=}g(T){=}0$ and $f(T){=}g(0){=}1$. 
This squeezes the qubit into a slightly shorter chain, as the boundary spin is now in a product state with the remainder of the chain. Note that the time-dependent part of the Hamiltonian in Eq.~\eqref{eq:singlequbitdyn} is $D_2$-invariant, and so preserves the degeneracy of the ground state.

To determine the effect of the single-qubit dynamics, we make use of two conserved quantites: $\Sigma_n^\z$ and $\Sigma_n^\x$. The former is clearly conserved; to see that the latter is, too, note that $[(S^\z)^2,\exp({i\pi S^\x})]=0$ and in particular, $\exp(i\pi S^\x)\ket{S^\z{=}0}=-\ket{S^\z{=}0}$. Now imagine the qubit starts in a $+1$ eigenstate of $\Sigma_n^\z$, i.e.\ the state 
\begin{equation}
  \ket{\psi(0)}=\ket{0}_n\equiv\ket{\Sigma_n^\z{=}{+}1,H_{n}^{\rm term}=0}\,. 
\end{equation}  
After the adiabatic dynamics it becomes 
\begin{equation}
  \ket{\psi(T)}=\ket{\Sigma_n^\z{=}{+}1,(S_1^\z)^2{=}0,H_{n{-}1}^{\rm term}{=}0}\,.
\end{equation}  
But due to the product form of $\Sigma_n^\z$, this is nothing other than the state $\ket{S^\z{=}0}\otimes \ket{\Sigma_{n{-}1}^\z{=}{+}1,H_{n{-}1}^{\rm term}{=}0}$, meaning $\ket{\psi(T)}=\ket{S^\z{=}0}\otimes\ket{{0}}_{n{-}1}$, up to some unknown phase. By the same argumentation, $\ket{{1}}_n$ becomes $\ket{S^\z{=}0}\otimes\ket{{1}}_{n{-}1}$, up to a possibly different phase. 

Up to an irrelevant global phase, the effect of the dynamics is a rotation of the qubit around the $\z$-axis, the amount depending on the relative phase accumulated by the two states $\ket{0}_n$ and $\ket{1}_n$. The relative phase is fixed by the other conserved quantity. 
Consider the time evolution of a logical qubit initialized in the $+1$ eigenstate of $\Sigma_n^\x$, $\ket{+}_n$. Because $\exp(i\pi S^\x)\ket{S^\z{=}0}=-\ket{S^\z{=}0}$, the eigenvalue of $\Sigma_{n{-}1}^\x$ in the final step must be $-1$, so $\ket{+}_n$ is transformed into $\ket{S^\z{=}0}\otimes\ket{-}_{n{-}1}$. Therefore, the dynamics effects a $\pi$ rotation of the qubit about the $\z$ axis. 

The dynamics can just as well be run in reverse, meaning we can perform any single qubit operation by first uncoupling and then recoupling the boundary spin. Aligning the local field to $\m$ during the forward stage and to $\m'$ during the reverse results in a qubit rotation of $2\cos^{-1}(\m\cdot\m')$ around the axis $\m\times\m'$.  The local field rotation on the boundary spin should also be done adiabatically with respect to the local gap which we assume is $\sim\Delta$ ; see Fig.~\ref{fig:holo}.   

Two-qubit operations can be similarly realized by appropriately coupling the ends of two neighboring chains $A$ and $B$ while decoupling them from their respective chains. For a judicious choice of coupling, this results in a {\sc cphase} gate followed by a joint $\pi$ rotation about the $\widehat{x}$ axis. The Hamiltonian for this process, $H^{AB}$ consists of the time-independent part $H^A_{{n{-}1}}+H^B_{{n{-}1}}$ and a time-dependent part 
\begin{align}
\label{eq:cphase}
H^{AB}(t)=f(t)J\,W^{AB}+g(t)J(\vec{S}_1^A\,{\cdot}\,\vec{S}_{2}^A+\vec{S}_1^B\,{\cdot}\,\vec{S}_{2}^B),
\end{align} 
where the interaction term is given by 
\begin{equation}
  W^{AB}=[(S_1^\x)^2-(S_1^\y)^2]^A\otimes [S_1^\z]^B+[S_1^\z]^A\otimes [(S_1^\x)^2-(S_1^\y)^2]^B\,.
\end{equation}
The ground state of this interaction term is 
\begin{equation}
  \ket{\xi}^{AB}=\tfrac 12({-}\ket{xx}+\ket{xy}+\ket{yx}+\ket{yy})^{AB}\,,
\end{equation}  
for $\ket{j}\equiv\ket{S^j=0}$. 

\begin{figure}[hbtp!]
\begin{center}
\includegraphics{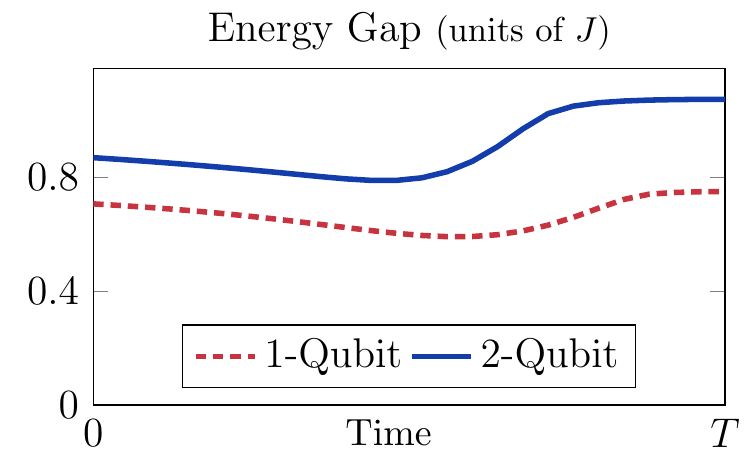}
\end{center}
\vspace{-5mm}
\caption{\label{fig:gap}Energy gap to the first excited state during the dynamics of single and two-qubit operations. These results were obtained by exact sparse-matrix methods for chains of length ten for single qubit operation, and 5 and 6 for the two-qubit operation.}
\end{figure}

Again the argument is based on various conserved quantities  that arise  from symmetries of the interaction, which are shown in Table~\ref{tab:stab}. The Haldane phase of two chains is not known to be protected by these symmetries, but numerical calculation on chains of moderate length confirms that the degeneracy is indeed maintained. This is implicitly shown in Fig.~\ref{fig:gap}, which also depicts the gap to the excited states.

\begin{table}[!htbp]
\begin{center}
{\renewcommand{\arraystretch}{1.5}
\begin{tabular}{ccc}
$W$ symmetry & Eigenvalue  of $|\xi\rangle$  & Conserved quantity\\\hline
$R^\z\otimes \mathbbm{1}$ & $-1$ & $\Sigma^\z\otimes \mathbbm{1}$\\
$\mathbbm{1} \otimes R_{\hat{z}}$ & $-1$  & $\mathbbm{1}\otimes \Sigma^\z$\\
$R^\u\otimes R^\u$ & $\phantom{-}1$  & $\Sigma^\u\otimes \Sigma^\u$\\
$R^\v\otimes R^\v$ & $\phantom{-}1$  & $\Sigma^\v\otimes \Sigma^\v$\\
$\sqrt{R^\z}\otimes R^\x$ & $-$i & $\sqrt{\Sigma^\z}\otimes \Sigma^\x$
\end{tabular}
}
\end{center}
\caption{\label{tab:stab}
Symmetry operators of the interaction $W^{AB}$ of Eq.~\ref{eq:cphase}, their eigenvalues in its groundstate, and the corresponding conserved quantities for two chains under the associated dynamics. The first factor in the tensor product acts on system $A$, the second on $B$. The conserved quantities fix the action on the encoded qubits to be the {\sc cphase} gate (followed by local rotations). The symmetry operators are all rotations, defined by $R^\m=\exp(-i\pi S^\m)$ and $\sqrt{R^\m}= \exp(-i\frac{\pi}{2} S^\m)$, with $\u=\tfrac{1}{\sqrt{2}}(\x+\y)$ and $\v=\tfrac{1}{\sqrt{2}}(\x-\y)$. Because the terms besides $W^{AB}$ in the Hamiltonian are rotationally-invariant, applying these rotations to their entire respective chains leads to the listed conserved quantities.}
\end{table}

To see that this operation implements a two-qubit unitary gate, we group the set of conserved quantities according to the eigenvalue of $\ket{\xi}$ and start with the first two, corresponding to $-1$. The joint eigenstates are product encoded states, and it is immediately clear, using the same analysis as the single qubit case, that the action of the dynamics is given by an operator of the form 
\begin{align}
U^{AB}=\begin{pmatrix} 0 & 0 & 0 & \alpha\\ 0 & 0 & \beta & 0\\ 0 & \gamma & 0 & 0 \\ \delta & 0 & 0 & 0\end{pmatrix},
\end{align}
where $\alpha,\beta,\gamma,\delta$ are complex numbers of unit magnitude. Now we consider the second pair of conserved quantities and determine that their joint eigenstates are the (unnormalized) states $\ket{01}\pm\ket{10}$ and $\ket{00}\pm i\ket{11}$. Working out the action of $U^{AB}$ on these states and using the overall $+1$ eigenvalue of $\ket{\xi}$ fixes $\delta=-\alpha$ and $\gamma=\beta$. Finally, the last conserved quantity has a nondegenerate spectrum, and so does not require a partner to determine a basis for the encoded states from its eigenstates, which happens to be the canonical Bell states. Again applying $U^{AB}$ and using the overall eigenvalue $-i$ of $\ket{\xi}$ as before gives $\alpha=-\beta$. Thus,
\begin{align}
U^{AB}=\begin{pmatrix} 0 & 0 & 0 & -1\\ 0 & 0 & 1 & 0\\ 0 & 1 & 0 & 0 \\ 1 & 0 & 0 & 0\end{pmatrix}
\end{align}
up to an irrelevant overall phase, which is nothing other than $(\bar{\sigma}^{\hat{x}}\otimes \bar{\sigma}^{\hat{x}})$\textsc{cphase}. 

\subsection{Initialization and Readout}

To initialize and measure the encoded quantum information, we may appeal to a method developed for a measurement-based precursor to the present scheme~\cite{miyake_quantum_2010}. Here we again adiabatically turn off the boundary spin coupling, but now do not turn on any local field. Instead, after the coupling is off, the boundary spin is measured in the basis $\ket{S_z=m}$. A result $m=1$ ($m=-1$) corresponds to a projection of the qubit onto $\ket{0}$ ($\ket{1}$), while $m=0$ corresponds to a $\pi$ rotation around the $\hat{z}$ axis. This can be understood as a consequence of addition of angular momentum; since the initial state has \spinhalf\ and the dynamics is adiabatic and preserves rotational invariance, the output state also has \spinhalf. But it is now more naturally expressed as a combination of spin-1 and \spinhalf, and by using the Clebsch-Gordan coefficients for this case we can quickly deduce the effect of measuring the spin-1 system. 

Once the system is in the ground state, encoded qubits may be initialized via this procedure by simply measuring until an $|m|=1$ outcome is obtained. This produces either $\ket{0}_n$ or $\ket{1}_n$ and the computation can begin. The problem of initialization is therefore reduced to preparing the system in the ground state, which can be done either by preparing a ground state at some point in the phase convenient to do so, e.g.\ the AKLT point, and then slowly altering the Hamiltonian to any other point in the Haldane phase which is convenient to implement,  e.g.\ the Heisenberg antiferromagnet 
or by actively cooling the state at a fixed point in the phase. For instance, Ref.~\cite{kraus_preparation_2008} makes use of the frustration-free property of the AKLT state to construct a Liouvillian map acting on neighboring pairs of spins and a local Markvoian environment whose output converges to the AKLT ground subspace in a time which scales linearly with the chain length.

The procedure of~\cite{miyake_quantum_2010} may also be used to read out the encoded qubit state after the computation is complete. This process is indeterministic due to the $m=0$ outcome, which can be dealt with in one of two ways. First, for outcome $m=0$, the boundary spin can be recoupled (after turning on a local field $S_z^2$) and the measurement attempted again. Second, controlled-{\sc not} gates to several other encoded qubits could first be performed, and then the encoded qubits measured simultaneously, taking as the outcome the majority of nonzero results.

\section{Noise Analysis and Error Correction}
\label{NoiseAnalysis}

Relative to a `bare' encoding of qubits in individual spins, the extra overhead in our scheme brings with it several advantages. 
Some types of errors are avoided entirely, while some others are suppressed.  As already mentioned, the encoding is immune to noise and disorder which respect the symmetry and which do not destroy the SPTO of the system; that is, the degeneracy of the ground space is protected.  This helps to avoid dephasing. The holonomic nature of the gates makes them inherently resistant to timing errors and intensity fluctuations. Additionally, logical gates are also resistant to spurious fields generated during the dynamics but having the symmetry appropriate to the gate being performed. This is especially appealing because the logical gates can be implemented by dynamically turning on and off control fields having fixed orientation: Errors in the field directions are then unknown quenched (systematic) errors which can be made arbitrarily small by composite pulse sequences~\cite{brown_arbitrarily_2004}.

Apart from initialization and readout, the logical operations maintain the degeneracy and gap of the energy spectrum, so we may hope that both leakage and logical errors can be suppressed by operating the system at low temperatures.  A full thermal stability analysis is beyond the scope of this paper, but a simple analysis presented in Sec.~\ref{GapProtection} shows that single-site noise does not translate into logical error without an energy penalty~\cite{refael_private_2009}. Indeed, only single-site noise on the boundary affects the logical information, and as such, noise in the bulk can be dealt with by cooling the chain. Rotations of the bulk spin carry an energy cost increasing with the amount of rotation; when using the encoded qubits in an active error-correction scheme, these rotation errors are digitized, and small rotations only lead to digitized error with small probability. Thus, the error rate decreases with decreasing energy of the noise. This limited protection against single-site noise unfortunately does not extend to two sites, a fact which is to be expected as the gate operations themselves involve only two sites. 

Although the SPTO of our systems is not generically robust to perturbations which do not have the $D_2$ symmetry, numerical simulations on the Haldane phase have shown a robustness of this phase to homogenous local fields.  Specifically, prior work shows that the phase is maintained for magnetic field perturbations of arbitrary direction affecting the entire chain homogeneously, up to magnitude roughly equal to the excitation gap~\cite{fath_solitonic_1993,fath_massless_1998,miyazaki_spin-reorientation_2006}.
On the other hand, direct numerical calculation shows that the ground state splitting is roughly proportional to the local field strength for a field acting only on the boundary spin at the AKLT point, as would be expected for a bare qubit implementation.

\subsection{Symmetry and gap protection}
\label{GapProtection}

Because we encode one qubit into a many body chain there are several possible locations for error that must be accounted for.  A complete summary of error mechanisms and the effect on encoded quantum information is displayed in Table \ref{tab:errors}.  The main effects are either to produce an error on the encoded information directly or to couple to states outside the qubit subspace which we denote leakage error.

\begin{table}[htdp]
\begin{center}
$\phantom{blahbla}$
{\renewcommand{\arraystretch}{1.5}
\setlength{\tabcolsep}{20pt}
{\small
\begin{tabular}{ccc}
\multicolumn{2}{c}{Error type}  & Effect \\
\hline
\multirow{3}{*}{Memory} & $D_2$-invariant & Logically protected \\
& Bulk & $p_L=0$, $p_{\ell}\sim \left(\frac{||h||}{\Delta}\right)^2$\\
& Boundary & $p_L \sim \frac{||h||}{\Delta}$ \\\hline
\multirow{3}{*}{Gate} & $D_2$-invariant &  Logically protected \\
& Quenched  & Systematic $p_L\sim \frac{||h||}{\Delta}$ \\
& Stochastic & $p_{L}\sim \frac{||h||}{\Delta}$
\end{tabular}
}
}
\caption{Error mechanisms and effects on encoded quantum information.  Here $p_{L(\ell)}$ are the logical(leakage) error probabilities, $h$ is the perturbation, either to the system Hamiltonian (Memory) or the gate Hamiltonian (Gate), and $||\cdot ||$ is the operator norm.  The Haldane gap is $\Delta$ and we assume the time to perform a gate is $O(1/\Delta)$.  Relatively benign are the systematic errors can be corrected using composite pulse sequences of length $k$ yielding an effective logical error $p_L=O((||h||/\Delta)^k)$~\cite{brown_arbitrarily_2004}, and leakage errors which are correctable by cooling. Other noise mechanisms yielding logical errors must be handled using quantum error correction.}
\end{center}
\label{tab:errors}
\end{table}%

Note that bulk and boundary perturbations of the system produce very different behavior, due to the fact that noise in the bulk produces local excitations, incapable of immediately causing logical errors as they cannot distinguish the logical states.  Provided bulk excitations are cooled at a rate faster than the dispersion, they will not propagate to the boundaries to cause logical errors.  In contrast, the boundary spin-$1$ particles are effectively free \spinhalf\ degrees of freedom susceptible to local energy shifts which translate into logical errors, albeit with an energy cost.

To gain more insight into why the bulk errors are relatively benign, consider the valence-bond solid as a caricature of the Haldane phase ground state (which is exact at the AKLT point)~\cite{affleck_valence_1988}.  A half-infinite chain has the following form in the Schwinger representation~\cite{arovas_extended_1988}:
\begin{align}   
\ket{\textsc{vbs}}=\left(\alpha a_0^\dagger+\beta b_0^\dagger\right)\prod_{j\geq 1}\left(a_j^\dagger b_{j{+}1}^\dagger-b_j^\dagger a_{j{+}1}^\dagger\right)\ket{\text{vac}},
\end{align}
where $a^\dagger_j$ and $b^\dagger_j$ are harmonic oscillator creation operators for modes $a$ and $b$ at site $j$ 
with the constraint $a^\dagger a + b^\dagger b = 2$ for every $j$,
while $\alpha$ and $\beta$ are complex coefficients for basis states of the edge mode. Single spin rotations mix the creation operators linearly, and in particular a $\hat{z}$-axis rotation by $\theta$ maps $a^\dagger$ to $a^\dagger e^{i\theta/2}$ and $b^\dagger$ to $b^\dagger e^{-i\theta/2}$. Letting $C^\dagger_{j,j{+}1}= \left(a_j^\dagger b_{j{+}1}^\dagger-b_j^\dagger a_{j{+}1}^\dagger\right)$, it is easy to work out that a rotation $R_j$ of site $j$ produces the transformation
\begin{align}
\label{eq:rotation}
R_jC^\dagger_{j,j{+}1}R_j^\dagger\!=\!\cos\tfrac{\theta}{2}C^\dagger_{j,j{+}1}{-}i\sin\tfrac{\theta}{2}\!\left(a^\dagger_jb^\dagger_{j{+}1}{+}b^\dagger_ja^\dagger_{j{+}1}\right).
\end{align}

The second term produces an excited state from the vacuum, and thus rotation of a site in the bulk of the chain leaves the encoded qubit unaffected while producing a linear superposition of ground and excited states. Note that the weights in the superposition depend on the amount of rotation. However, since the superposition does not depend on the encoded information, it can therefore be corrected without damaging the qubit, e.g.\ by cooling. On the other hand, a rotation of the boundary spin produces a superposition between the original encoded qubit in the ground state and a rotated version in the excited state. Again, the amount of rotation determines the amplitude of the excited state, and so larger rotations cost more energy.

To estimate the leakage probability, note that the excited states are split from the ground states by the Haldane gap.  For a local perturbation $h$ acting in the bulk the amplitude of coupling to the excited states is $\sim ||h|| T_{\rm gate}$ and since the three step holonomic gate time is $T_{\rm gate}= 3/\Delta_{\rm min}=O(1/\Delta)$, where $\Delta_{\rm min}$ is the minimum gap during each step, then the probability to leak into the excited states is $p_{\ell}\sim (||h||/\Delta)^2$.
On the other hand, the edge modes are shifted in energy by a local perturbation without gap protection implying a logical error over a gate time scaling like $p_L\sim ||h||/\Delta$.

\begin{figure}[ht]
\begin{center}
\includegraphics[width=8cm]{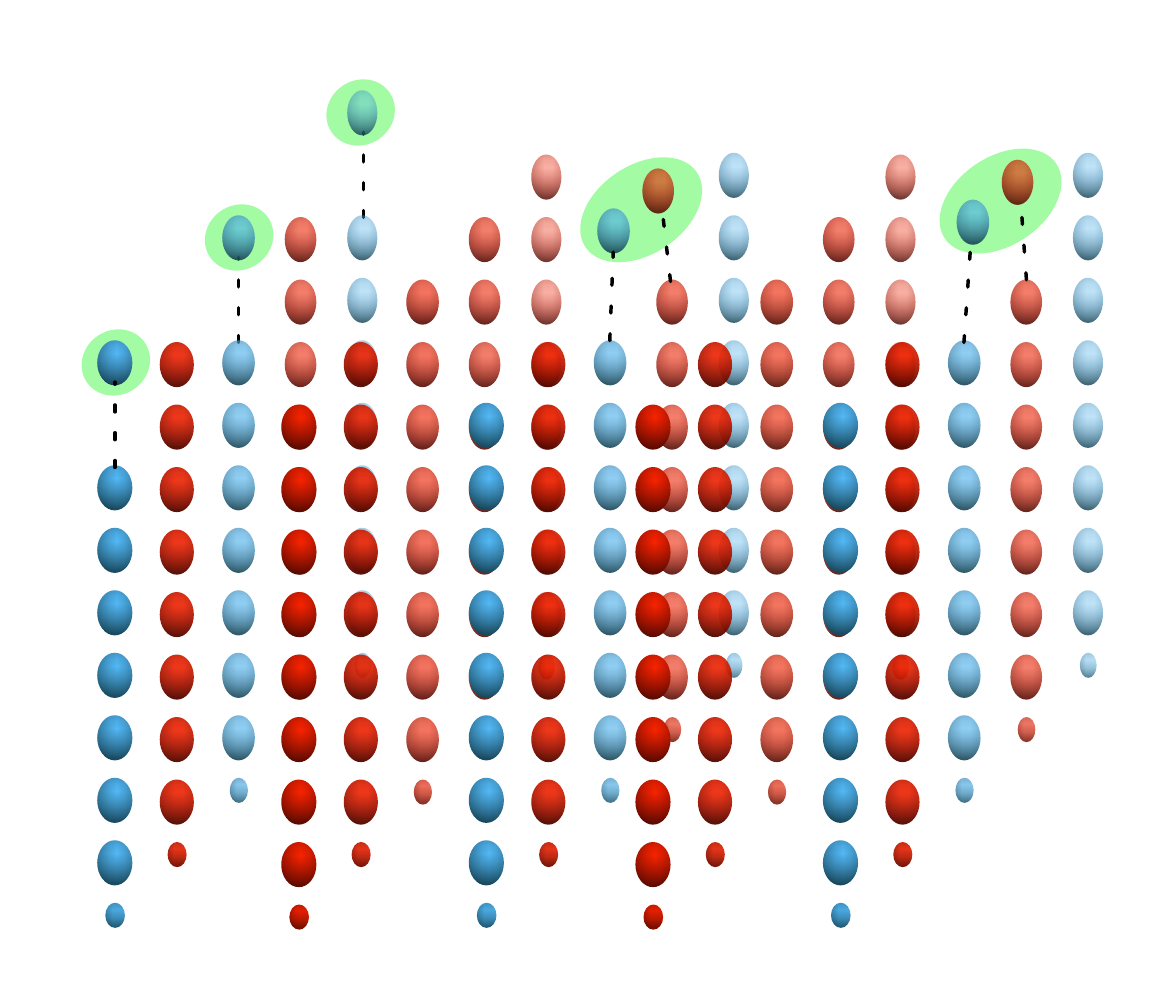}
\caption{\label{fig:3Darch}  {A potential architecture for fault tolerant computing using an array of vertical Haldane chains.  Each short Haldane chain consists of 8 bulk spin$-1$ particles (big dots) and one boundary spin$-1/2$ (small dots), and has a degenerate ground subspace which defines a physical qubit.  The $3\times 3$ sub array of blue chains are data qubits and the red chains in between are ancilla qubits which provide room for fault tolerant swapping between data qubits and for rounds of error diagnosis and correction.} Entropy can be dumped into the environment by resetting the ancilla, either by measurement or cooling.  Qubits are measured by adiabatically decoupling one boundary spin, measuring that spin and if a result of $S^z=\pm 1$ is obtained succeeding, otherwise recoupling the boundary spin to the chain and trying again.  The figure illustrates a logical qubit encoded in a 9 qubit Bacon-Shor quantum error correction code.  A transversal logical X gate is depicted on the left column with the green ovals representing a gate acting on the boundary spins that have been adiabatically dragged away from their Haldane chains.   Two {\sc cphase} gates between level 0 data and ancilla qubits are also depicted.}
\end{center}
\end{figure}

\subsection{Quantum Error Correction}

While the hardware protected quantum gates we have described reduce error rates due to environmental noise and control error,  software based quantum error correction is needed to achieve fully fault tolerant quantum computation.   In some 
ways our gate mechanism constrains the type of architecture and choice of quantum error correction code (QECC).  First, because we rely on adiabatically turning interactions on and off, our architecture is likely to allow only nearest neighbor interactions between qubits.  It may be possible that a nonlocal coupling could be engineered using, for example, an optical mode in a fibre~\cite{serafini_distributed_2006}; however, such interactions are typically weak and could prove difficult to wire in a scalable system.  Second, since our qubits are degenerate by design, there is no bias towards any particular local Pauli errors.  This is unlike the situation with many physical realisations of qubits which are non degenerate, such as hyperfine split ground states of trapped ions or atoms, superconducting phase qubits, etc, which are inherently more resiliant against bit flip errors in the energy basis versus phase errors.  This being the case, it is reasonable to chose quantum error correction codes, and a concatenation method which is unbiased toward $X$ or $Z$ error.   Some particular QECCs and architectures are well suited to this very situation~\cite{gottesman_introduction_2009}.    

One suitable code is the Bacon-Shor 9 qubit QECC embedded in a 2D spatial architecture.  It was shown in Ref.~\cite{spedalieri_latency_2008} that ``padding" a 2D array with ancilla qubits in between data qubit provides sufficient room to perform fault tolerant swapping of information even when restricted to nearest neighbor interactions.  Assuming an adversarial local error model and equal error rates for memory and gate errors, the threshold for fault-tolerant computing in that architecture is $p_{th}=1.3\times 10^{-5}$.  This can be improved using other software style strategies such as concatenated dynamical decoupling pulses~\cite{west_high_2010}. These thresholds are about a factor of 10 worse than a nonlocally corrected architecture.  The Bacon-Shor code has the advantage of low latency (i.e.\ overall faster performance) relative to other CSS codes such as the 7 qubit QECC and does not require cat state verification of ancilla.  Furthermore, this code also accommodates measurement-free QEC with only a small reduction in the threshold~\cite{paz-silva_fault_2010} which would obviate the need for fitting high efficiency detectors within the lattice of chains.

An illustration of an architecture using a 2D array of parallel chains of encoded qubits is shown in Fig.~\ref{fig:3Darch}. Here the chains are depicted with real \spinhalf\ boundary spins, which could however be removed with some minor changes to the protocol.  As stated above, except at the AKLT point, the ground state manifold suffers a small degeneracy splitting which decreases exponentially with the chain length.  The splitting translates into a small timing error in logical gates and opens up a zero energy leakage channel to the other edge mode.  Both could be corrected with QEC, but another possibility is to use both edge modes as qubits and process them by manipulating both ends of the chains. 

Because each qubit is a collective degree of freedom of a spin chain, it might be thought that a local error model is not appropriate.  However, all the chains are separated in space far enough to have negligible interactions between chains.  Any correlated errors within a spin chain effectively act as either a leakage error when they occur in the bulk of the chain which can then propagate to the boundaries to create a logical error, or directly as a logical error at the boundary.  

Leakage errors  require special care during error correction because once such an error occurs in one chain, subsequent gates between chains that are even perfect with the logical subspace can propagate error in the leakage subspace.  This has shown not to be disastrous, however, provided one can perform leakage reduction at the lowest level of concatenation (i.e. at the level of  the physical chains)~\cite{aliferis_fault-tolerant_2007}.  An appealing hardware strategy for leakage reduction is to bathe the spin chains in a cold environment that removes excitations without requiring active monitoring of the system. Such a procedure does not restore the quantum information, but rather reduces leakage errors to standard logical qubit errors which can be corrected using quantum error correction.

\section{Discussion}
\label{Sum}

We have presented a new architecture for quantum computation in which the primitive information carriers are degenerate ground states of strongly correlated spin chains and gates are implemented by adiabatic holonomies.  
The hardware requirements of our model are fairly modest, as its information processing properties arise from it being a $D_2$ SPTO system with a degenerate ground state, not specific parameters in the Hamiltonian. Such SPTO spin chains can be realized using two-body nearest-neighbor couplings, as for example in the spin-1 Heisenberg antiferromagnet.  Spatial variations in the precise nature of the couplings are allowed, provided they maintain the symmetry, making the system robust to modest disorder. Because the gate operations occur on the boundary, the couplings in the bulk of the chain can be fixed. 
By making use of strong few body interactions in the encoding of each logical qubit  (but not between logical qubits), the symmetry protected holonomic mechanism reduces some memory error and logical gate error rates without introducing new error channels outside the assumptions of the standard fault tolerant threshold theorems~\cite{gottesman_introduction_2009}.  

This hardware assisted approach should make it easier to reach the thresholds needed for fault tolerant quantum computation.  We envisage a variety of potential implementations for this scheme, including ultracold polar molecules~\cite{brennen_designing_2007}, trapped atoms~\cite{trotzky_controlling_2010} or quantum dots~\cite{willems_van_beveren_spin_2005}. 

While our symmetry protected scheme enjoys many benefits it is, as we have pointed out, subject to logical errors due to local non-$D_2$ symmetric perturbations.  That being the case one may ask whether it would not be simpler and equally advantageous to simply encode each logical qubit in a degenerate subspace of a single particle, e.g.\ a spin-$1$ particle subject to a local Hamiltonian $H_1=-\Delta (S^z)^2$ which provides for a gap $\Delta$ to logical errors, and to perform holonomic gates on those degenerate degrees of freedom.   A universal holonomic gate set with such an encoding seems difficult \footnote{A direct approach to find holonomic single logical qubit gates with this encoding seems to only work for rotations about one axis.}, and notwithstanding, would not be as robust for the following reasons.  First, consider how one protects the degeneracy which defines the qubit encoding.  In our spin chain scheme, the encoding is protected by a phase which is robust to a large family of perturbing Hamiltonians, i.e.\ those which preserve the discrete symmetry group $D_2$. The single-particle implementation would need to be protected against perturbations generating a continuous Lie group and any unwanted rotation in the direction of the field in $H_1$ would generate a logical error.  Quite generally, nature abhors a symmetry and we make use of many body interactions in order to provide for it, in a way that single particle interactions cannot.  Second, one needs to provide for protected gates to process the quantum information.  In the spin chain model the gates are performed by manipulating the boundary spins and are fully protected by the SPTO.  In a single particle encoding any holonomic gate would require much more restrictive control of the Hamiltonian because there is less freedom in the allowed perturbations.

We conclude with some potential directions for future work.  First, one could ask whether a similar scheme is possible in chains of spin-1/2 particles, rather than spin-1.  While it is possible to construct a spin-1/2 Hamiltonian in the Haldane phase by using alternating ferromagnetic-antiferromagnetic couplings~\cite{Hida_Haldane_1992}, a direct mapping of the two-body couplings required for two-qubit gates presented here will lead to three-body couplings in the spin-1/2 model.  More exotic SPTO systems of spin-1/2 chains may provide alternative schemes with only two-body interactions.

Another natural question is whether a genuine 2D SPTO system~\cite{chen_2dSPTO_2011} can be used for such a holonomic scheme that protects both single- and two-qubit gates explicitly.  Our scheme constructed from 1D chains produces well-defined qubits on the ends; a 2D SPTO system is characterised by gapless 1D boundary systems but {without naturally defined qubits. The symmetry associated with this mode is a highly nonlocal operator and as such would be more resilient to local errors on the boundary.  It would be worthy to investigate whether quantum information could indeed be encoded and be manipulated using nonlocal operators akin to string operations used as logical operations in surface codes and in the 2D quantum compass model.}

\section*{Acknowledgements}
\addcontentsline{toc}{section}{Acknowledgements}
We thank D.~Bacon, A.C.~Doherty, S.T.~Flammia, H.~Katsura, A.~Micheli, and G.~Rafael for enlightening discussions. JMR and AM acknowledge the support of the Perimeter Institute, which receives funding from the Government of Canada through Industry Canada and Ontario-MRI.  SDB and GKB acknowledge support from the ARC through the ARC Centre of Excellence in Engineered Quantum Systems (EQuS), project number CE110001013. JMR acknowledges support from the Swiss National Science Foundation (through the National
Centre of Competence in Research `Quantum Science and Technology' and grant
No. 200020-135048) and by the European Research Council (grant 258932). AM acknowledges support from the EU project (QCS). GKB received support from the EC project AQUTE. The research leading to these results has received funding from the European Community's Seventh Framework Programme (FP/2007-2013) under grant agreement no.\ 247687 (Integrating Project AQUTE).



\end{document}